\documentclass[12pt]{article}

\begin{document}
\input {epsf}

\newcommand{\beq}{\begin{equation}}
\newcommand{\eeq}{\end{equation}}
\newcommand{\beqa}{\begin{eqnarray}}
\newcommand{\eeqa}{\end{eqnarray}}

\def\ov{\overline}
\def\onlyif{\rightarrow}

\def\openone{\leavevmode\hbox{\small1\kern-3.8pt\normalsize1}}

\def\a{\alpha}
\def\b{\beta}
\def\g{\gamma}
\def\r{\rho}
\def\minus{\,-\,}
\def\eks{\bf x}
\def\kay{\bf k}

\def\ket#1{|\,#1\,\rangle}
\def\bra#1{\langle\, #1\,|}
\def\braket#1#2{\langle\, #1\,|\,#2\,\rangle}
\def\proj#1#2{\ket{#1}\bra{#2}}
\def\expect#1{\langle\, #1\, \rangle}
\def\trialexpect#1{\expect#1_{\rm trial}}
\def\ensemblexpect#1{\expect#1_{\rm ensemble}}
\def\kpsi{\ket{\psi}}
\def\kphi{\ket{\phi}}
\def\bpsi{\bra{\psi}}
\def\bphi{\bra{\phi}}

\def\ditto{\rule[0.5ex]{2cm}{.4pt}\enspace}
\def\th{\thinspace}
\def\ni{\noindent}
\def\thirty{\hbox to \hsize{\hfill\rule[5pt]{2.5cm}{0.5pt}\hfill}}

\def\set#1{\{ #1\}}
\def\setbuilder#1#2{\{ #1:\; #2\}}
\def\Prob#1{{\rm Prob}(#1)}
\def\pair#1#2{\langle #1,#2\rangle}
\def\Id{\bf 1}

\def\dee#1#2{\frac{\partial #1}{\partial #2}}
\def\deetwo#1#2{\frac{\partial\,^2 #1}{\partial #2^2}}
\def\deethree#1#2{\frac{\partial\,^3 #1}{\partial #2^3}}

\def\openone{\leavevmode\hbox{\small1\kern-3.8pt\normalsize1}}

\title{Eavesdropping without quantum memory}
\author{H. Bechmann-Pasquinucci\\
\small
{\it University of Pavia,  Dipartimento di Fisica 
"A. 
Volta", via Bassi 6, I-27100 Pavia, Italy}\\
\small
{\rm UCCI.IT}, {\it via Olmo 26, I-23888 Rovagnate, Italy}}
\date{March 31, 2005}
\maketitle

\abstract{In quantum cryptography the optimal eavesdropping strategy 
requires that the eavesdropper uses quantum memories in order to optimize 
her information. What happens if the eavesdropper has no quantum memory?  
It is shown that the best strategy is actually to adopt the simple 
intercept/resend strategy.

} \vspace{1 cm} \normalsize

\section{Introduction}
With the development of quantum information theory, traditional quantum
state discrimination has, in many cases, been given a twist. It is no 
longer
just the simple question of identifying one state drawn from a known set
of states. Often there is additional information available after the
interaction with the 'unknown' system or even after the measurement has
been performed. For 
example, in the BB84 protocol \cite{BB84} for quantum
cryptography \cite{review} the eavesdropper, Eve, knows that the quantum 
system 
is
prepared with equal probability in a states belonging to a set of states
made by two mutually unbiased bases. Moreover, she knows that after 
her
eavesdropping, i.e. after the interaction with the 'unknown' quantum state,
she will learn in which basis the system was originally prepared. She then 
uses this 
additional classical information, to gain more 
information about
the initial state.

For the BB84 protocol, the optimal eavesdropping strategy 
\cite{optBB84}, consists in
intercepting the system prepared by Alice, attach an ancilla and let the
combined system undergo a unitary interaction. After the interaction the
original system is forwarded to Bob, whereas Eve keeps the ancilla. In
this way she can transfer some of the information about the original state
to her ancilla, with the cost of disturbing the original state and hence
introduce errors on Bob's part. The more information Eve transfers to her
own system, the more she is disturbing the original system and the higher
error rate she is introducing. In order for Eve to get the maximum
information out of her ancilla, it is usually assumed that she does not
measure her ancilla until after the public discussion between Alice and
Bob. In this way she can use the knowledge that she gains by passively
listening to the public discussion to select the measurement best suited
for each ancilla. However, this requires that Eve is able to store her
ancilla for a certain amount of time in a quantum memory. Here we ask the
question, what happens if Eve does not have a quantum memory?

In this paper we consider the standard BB84 protocol \cite{BB84} for 
qubits and discuss basically two different scenarios: the simple 
intercept/resend eavesdropping \cite{intercept} and 
eavesdropping using an ancilla --- but 
without 
a quantum memory. In both cases we consider a range of von Neumann 
measurements.

The scenario which is considered here is very simple, but the underlying 
question is both important and interesting because it concerns not only 
eavesdropping, but a much more general scenario: What happens when state 
discrimination is combined with additional classical information? What is 
the 
optimal measurement, when there later will be given additional classical 
information? These are questions which are interesting to consider in 
full generality. The study made in this paper should be considered only 
the beginning.

\section{Intercept/resend eavesdropping}
Consider the BB84 protocol for qubits, which uses two mutually unbiased 
bases for 
the secret key creation. We assume that Alice and Bob use the $x$ and 
the $y$-basis, and use of the following 
definition of the states, 
\begin{eqnarray}
\ket{x_{\pm}}=\frac{1}{\sqrt{2}}(\ket{0} \pm \ket{1})~~~{\rm and} ~~~
 \ket{y_{\pm }}=\frac{1}{\sqrt{2}}(\ket{0} \pm i \ket{1})
\end{eqnarray} 
here expressed in the computational basis $\ket{0}$ and $\ket{1}$.

First we will consider intercept/resend eavesdropping, which historically
also was the first eavesdropping strategy to be considered. 
This strategy requires no quantum memories, and it is
reviewed in order to compare with the optimal eavesdropping strategy
{\it without} quantum memory. It consists very simply in Eve intercepting 
the
qubit prepared by Alice while in transit to Bob, she then estimates the
state of the qubit by means of a measurement, and prepares a new qubit in
the state that she found and sends it to Bob. 
\begin{figure}[t]
\begin{center}
\begin{picture}(100,100)(0,0)
\put(50,50){\vector(1,0){40}}
\put(50,50){\vector(-1,0){40}}
\put(50,50){\vector(0,1){40}}
\put(50,50){\vector(0,-1){40}}
\put(50,50){\vector(3,1){35}}
\put(50,50){\vector(-3,-1){35}}
\put(50,50){\vector(1,3){12}}
\put(50,50){\vector(-1,-3){12}}
\put(2,48){$-x$}
\put(48,5){$-y$}
\put(92,48){$+x$}
\put(48,92){$+y$}
\put(86,61){$+\phi$}
\put(61,87){$+\phi'$}
\put(8,38){$-\phi$}
\put(36,10){$-\phi'$}
\end{picture}
\caption{Here is shown the $x$ and the $y$ basis and the two measurement 
bases used by Eve, $\phi$ and $\phi'$. Notice that the states are drawn on 
the equator of the Poincare sphere.} 
\label{fig:fig1}
\end{center}
\end{figure}
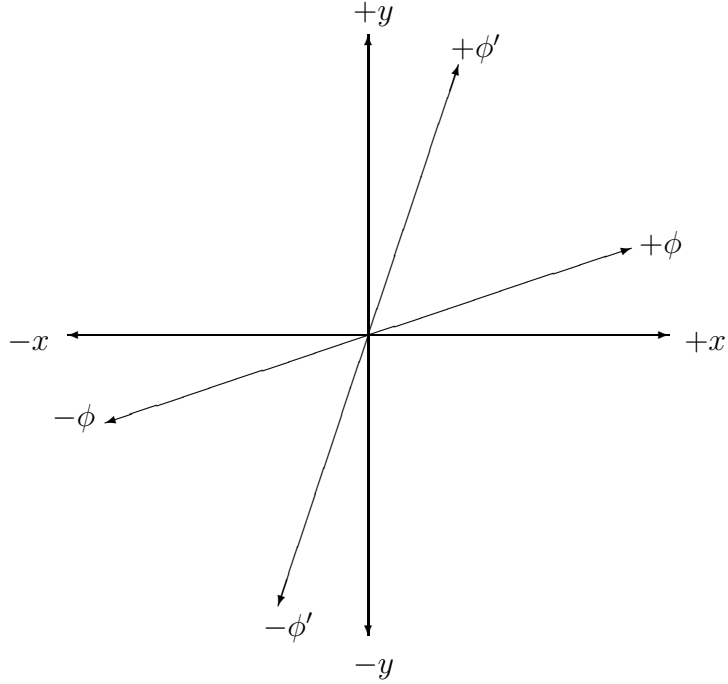
We assume that Eve performs a von Neumann measurement lying in 
the $xy$-plane\footnote{In higher dimension it will be necessary to 
consider POVMs \cite{POVM}}. 
It is possible to consider all measurement strategies of this 
kind in one go, by parameterizing the measurement as follows:
\begin{eqnarray}
\ket{+\phi}=\frac{1}{\sqrt{2}}(\ket{0}+{e}^{i\phi} \ket{1})~~~{\rm
and} ~~~\ket{-\phi}=\frac{1}{\sqrt{2}}(\ket{0}-{e}^{i\phi} \ket{1}),
\end{eqnarray}
where $\phi\in [ 0,\pi/4]$. Since this measurement is not symmetric with 
respect to the two bases, Eve 
will have different fidelities and disturbances in the two bases, moreover 
she will also introduce different error rates in the two bases. However, 
it 
is easy to restore the symmetry by letting Eve choose at random between two 
different measurements: namely the $\phi$-measurement and the measurement 
which corresponds to $\phi'=\pi/2 -\phi$ (see fig. 1). 
Notice that this symmetrization doesn't change Eve's average information.
When performing the $\phi$ or $\phi'$ measurement, Eve will obtain 
the 
following fidelities and disturbances in the two bases:
\begin{eqnarray}
& &F_{E,\phi}^{x}=F_{E,\phi'}^{y}=\frac{1}{2}(1+\cos\phi)~~~,~~~
D_{E,\phi}^{x}=D_{E,\phi'}^{y}=\frac{1}{2}(1-\cos\phi)
\nonumber \\
& &F_{E,\phi}^{y}=F_{E,\phi'}^{x}=\frac{1}{2}(1+\sin\phi)~~~,~~~
D_{E,\phi}^{y}=D_{E,\phi'}^{x}=\frac{1}{2}(1-\sin\phi)
\end{eqnarray}
Where as usual we have $F_{E,\phi}^{i}+D_{E,\phi}^{i}=1$.

Independently of whether Eve measures in the $\phi$ or the $\phi'$-basis, 
half of the times 
Alice has prepared the qubit in the $x$-basis, and half of the time 
she has prepared the qubit in the $y$-basis. Eve 
obtains the following amount of Shannon information \cite{ct}, 
respectively
\begin{eqnarray}
& &I_{E,\phi}^{x}=I_{E,\phi'}^{y}=1+F_{E,\phi}^{x}\log 
F_{E,\phi}^{x}+D_{E,\phi}^{x}\log D_{E,\phi}^{x} \nonumber \\
& &I_{E,\phi}^{y}=I_{E,\phi'}^{x}=1+F_{E,\phi}^{y}\log 
F_{E,\phi}^{y}+D_{E,\phi}^{y}\log D_{E,\phi}^{y}.
\end{eqnarray}
This means that Eve's average information is 
\begin{eqnarray}
I_E=\frac{1}{4}(I_{E,\phi}^{x}+I_{E,\phi}^{y}+I_{E,\phi'}^{x}+I_{E,\phi'}^{y}).
\end{eqnarray}
After her measurement Eve has to prepare a new qubit and send it to Bob. 
However, at this point in the protocol Eve doesn't know in which basis the 
original qubit was prepared. We consider the usual case where she prepares 
the same state that she found by her measurement and sends that to Bob.

Assuming that Eve measures in the $\phi$ basis, then the fidelity 
and disturbance which Bob finds, can be obtained by the
following argument: with probability $F_{E,\phi}^{i}$, $i=x,y$, Eve will
find the correct {\it guess} state and send it to Bob; where correct {\it
guess} state means that if Alice sent a $+$ state then Eve will identify 
the state as the  $+\phi$,  etc.
Assuming that Bob measures in the same basis as Alice, he will then  
have 
probability $F_{E,\phi}^{i}$ of obtaining the correct state. Whereas with 
probability $D_{E,\phi}^{i}$ Eve finds the wrong {\it guess} state and 
hence sends the wrong state to Bob. However, if Bob makes the wrong 
identification of the wrong state, he will actually obtain the correct 
state; this will happen with probability $D_{E,\phi}^{i}$. 
So in total Bob's probability for getting the correct 
state is:
\begin{eqnarray}
F_{B,\phi}^{x}=(F_{E,\phi}^{x})^2 
+(D_{E,\phi}^{x})^2=\frac{1}{2}+\frac{{\cos}^{2}\phi}{2} 
~~~(=F_{B,\phi'}^{y})\nonumber \\
 F_{B,\phi}^{y}=(F_{E,\phi}^{y})^2 
+(D_{E,\phi}^{y})^2=\frac{1}{2}+\frac{{\sin}^{2}\phi}{2}
~~~(=F_{B,\phi'}^{x})
\label{eq:fib}
\end{eqnarray}
As for Eve, due to symmetry we have $F_{B,\phi}^{x}=F_{B,\phi'}^{y}$ and 
$F_{B,\phi}^{y}=F_{B,\phi'}^{x}$.

Making use of the expressions of Eve's fidelity and disturbance, one 
finds that 
Bob's overall 
fidelity $F_{B,\phi}=\frac{1}{2}(F_{B,\phi}^{x}+F_{B,\phi}^{y})=3/4$, and 
disturbance $D_{B,\phi}=1-F_{B,\phi}=1/4$ is independent of the 
measurement 
performed by Eve. However, if Eve doesn't alternate between the two bases 
$\phi$ and $\phi'$, Bob will find different fidelities and disturbances in 
his two bases, see eq.(\ref{eq:fib}).

There are a couple of special values of $\phi$ which are worth considering
more explicitly, namely the case $\phi=0$ and $\phi=\pi/4$: The case where
$\phi=0$ and hence $\phi'=\pi/2$, corresponds the the situation where Eve
is measuring at random in the $x$ and the $y$-basis, hence using the same
bases as Alice and Bob. The information that Eve obtains in this situation
is so-called deterministic information, becuase when the three of them use
the same basis, Eve knows the secret bit, whereas if she measures in the
wrong basis she will no nothing about the bit value. On
average Eve gains $1/2$ a bit of information.

In the case where $\phi=\pi/4$, we have a very special situation. In this 
case 
the $\phi$-measurement and the $\phi'$-measurement 
coincides and Eve no longer needs to choose at random between two 
measurements, since this single measurement treats the two bases 
symmetrically. 
This particular attack in known as the intercept/resend 
attack in the intermediate basis \cite{intermediate}. 
It can be shown that that this measurement optimizes Eve's 
probability of guessing the state correctly {\it independently} of the 
basis. Which means that Eve obtains the same amount of information on each 
single bit. However, her information is no longer deterministic, but 
probabilistic, which means that she knows she bit with a certain 
probability (different from 1).
Even if this measurement strategy gives Eve less 
information ($I_E\approx 0.39$), than measuring in the same basis as 
Alice and Bob, it is 
an advantage for Eve, when it is taken into account that Alice and 
Bob later will go through classical error correction and privacy 
amplification \cite{intercept}. This is due to the fact that  
probabilistic information is more 
robust during this process than deterministic information.

As a curious point should be mention that the states 
corresponding to the intermediate states, also play an optimal role in the 
game of quantum state targeting\cite{RS,FA} and Bell inequalities 
\cite{BPG}.

\begin{figure}[t]
\begin{center}  
\leavevmode
\hbox{%
\epsfxsize=5in
\epsffile{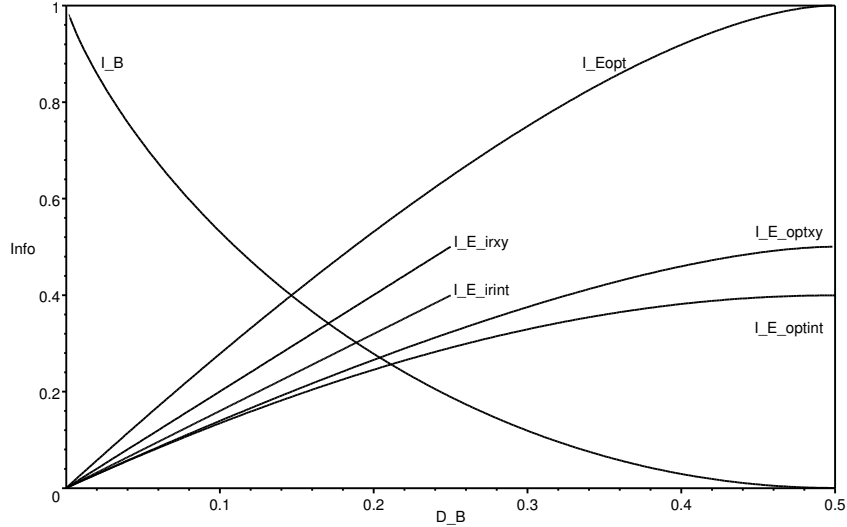}}
\caption{The information curves as a function of the disturbance detected 
by Bob: ${\rm I\_B}\equiv$ Bob's information. For Eve: 
${\rm I\_Eopt}\equiv$  optimal 
information {\it with} quantum memory, ${\rm I\_E\_irxy}\equiv$ 
intercept/resend 
measurement in the $x,y$ bases, ${\rm I\_E\_irint}\equiv$ 
intercept/resend 
measurement in intermediate basis, ${\rm I\_E\_optxy}\equiv$ optimal 
eavesdropping 
{\it without} quantum memory, measurement in the $x,y$ bases, ${\rm 
I\_E\_optint}\equiv$ 
optimal eavesdropping
{\it without} quantum memory, measurement in the intermediate basis. 
Notice that by changing $\phi\in [0,\pi/4]$, the information curve 
smoothly goes from 
${\rm I\_E\_{\ast xy}}$ to ${\rm I\_E\_{\ast int}}$ (see also Fig.3).} 
\end{center} 
\label{fig:fig2} 
\end{figure}

In order to compare with the results in the next section it is useful to 
display the information that Eve obtains as a function of the disturbance 
that she introduces. Eve can lower the disturbance by eavesdropping only on  
a fraction $f\in [0,1]$ of the transmitted qubits, where $f=0$ 
corresponds to no eavesdropping and $f=1$ to eavesdropping on all qubits. 
Assuming that eavesdropping is the only 
cause of errors, then the disturbance that Alice and Bob will find if Eve 
only eavesdrop on a fraction of the qubits is 
$D_{B}=f\cdot{D}_{B,\ast}=\frac{f}{4}$, since 
${D}_{B,\phi}={D}_{B,\phi'}=1/4$. 
Similarly, 
Eve's average information becomes 
$f\cdot I_{E}$. It is possible to express Eve's 
information in terms of 
the disturbance that she creates, since 
$f=4D_B$, which means $I_{E}(D_B) 
=4D_{B}I_{E}$. The corresponding information curves are 
displayed in figure 2. 
Notice that the information curves for intercept/resend eavesdropping are 
only defined 
up til the disturbance $D_{B}=1/4$, since this is the disturbance 
which 
Eve would introduce 
if she would eavesdrop on each single qubit. Furthermore it should be kept 
in mind that these information curves corresponds to the average 
information on the full key, since Eve obviously has no information when 
she doesn't eavesdrop.

Since $\phi \in [0,\pi/4]$, the two special cases considered above 
correspond to the end points of the interval, changing the parameter 
$\phi$ will therefore smoothly change the information curve from 
$I_{E,0}\equiv{\rm 
I\_E\_irxy}$ to 
$I_{E,\pi/4}\equiv{\rm I\_E\_irint}$, see Fig.2.

\begin{figure}[t]
\begin{center}  
\leavevmode
\hbox{%
\epsfxsize=5in
\epsffile{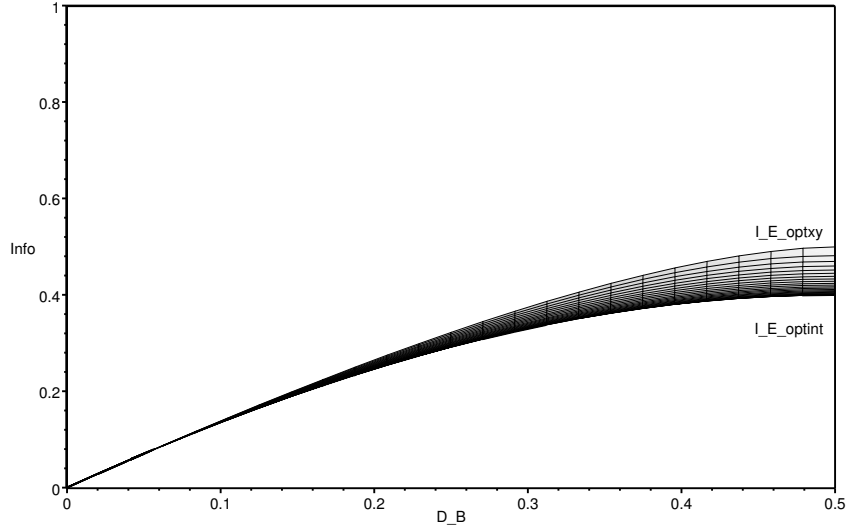}}
\caption{The bottom curve corresponds to measuring in the 
intermediate basis $(\phi=\pi/4)$ and the top curve corresponds to 
measuring 
in the $x$ or $y$ basis. 
} 
\end{center} 
\label{fig:fig3} 
\end{figure}

\section{Optimal eavesdropping}
The optimal eavesdropping strategy consists in Eve letting an ancilla
undergo a unitary interaction with the qubit prepared by Alice, after
which she sends on the (now disturbed) qubit to Bob and keeps her ancilla.
Eve usually stores her ancilla in a quantum memory and only 
performs a measurement on it
after she has learnt from the public discussion between Alice and Bob in
which basis the original qubit was prepared. In order to optimize her
information Eve has to measure her ancilla in same the basis as the qubit 
was
originally prepared.

When expressed in the computational basis, i.e. the $z$-basis, the optimal 
eavesdropping strategy can be written on the following simple, but 
asymmetric form 
\begin{eqnarray}
& &\ket{0}\ket{0}\stackrel{\bf U}{\longrightarrow}\ket{00}\nonumber\\
& &\ket{1}\ket{0}\stackrel{\bf U}\longrightarrow 
\cos{\alpha}\ket{10}+\sin{\alpha}\ket{01}
\label{eq:interaction}
\end{eqnarray}
where the lefthandside indicates the state before the interaction and the 
righthandside the state after the interaction of the qubit sent by Alice 
and the Eve's ancilla.

The fidelities of Bob and Eve are  
$F_{B}=(1+\cos\alpha)/2$, and 
$F_{E}=(1+\sin\alpha)/2$, respectively. 
The disturbance is $D_i=1-F_i$, 
where $i=\{ B, E\}$. 

It should be emphasized that with 
respect to the $x$ and $y$-basis the eavesdropping strategy is symmetric 
and the fidelities are therefore also the same for the two basis. The 
information curves for Bob and Eve $I_{i}=1+{D}_{i}{\log}_{2}{D}_{i}
+(1-{D}_{i}){\log}_{2}(1-{D}_{i})$, where $i=\{ B, E\} $  
are shown in figure 2, where Eve's disturbance has been expressed in 
terms of Bob's disturbance $D_B$, i.e. $D_{E}(D_{B})$. However, it should 
be remembered, that in order for Eve to optimize her fidelity and her 
information Eve has to perform her measurement
{\it after} the public discussion between Alice and Bob, which means 
storing her qubit in a quantum memory. 

We will now consider a situation which is less ideal for Eve --- but much 
more realistic as of today ---  namely where Eve has no possibility of 
storing her ancilla in a quantum memory and therefore has to make a
measurement right a way. As in the case of intercept/resend eavesdropping, 
Eve will still listen to the public discussion between
Alice and Bob, because even if the information about the original basis
preparation of the qubit arrives {\it after} she has performed her 
measurement she can still use the information to make an
interpretation of her measurement result and obtain more 
information. 

Consider again the situation where Eve measures with equal probability in 
either the $\phi$ or the $\phi'$ basis. Then, if she has let her ancilla 
undergo the interaction described in eq.(\ref{eq:interaction}), she will 
have the following measurement fidelity and disturbance:
\begin{eqnarray}
& &F_{E,\phi}^{x}=\frac{1}{2}(1+\cos\phi\sin\alpha)~~~,~~~
D_{E,\phi}^{x}=\frac{1}{2}(1-\cos\phi\sin\alpha)
\nonumber \\
& &F_{E,\phi}^{y}=\frac{1}{2}(1+\sin\phi\sin\alpha)~~~,~~~
D_{E,\phi}^{y}=\frac{1}{2}(1-\sin\phi\sin\alpha)
\end{eqnarray}
Where as usual we have $F_{E,\phi}^{i}+D_{E,\phi}^{i}=1$, and again 
due to symmetry (see figure 1) $D_{E,\phi}^{x}=D_{E,\phi'}^{y}$ 
and  $D_{E,\phi}^{y}=D_{E,\phi'}^{x}$. Bob's fidelity and disturbance 
do not change.

Based on the obtained fidelity and disturbance, the information can be 
computed  
$I_{E}(D_B)=\frac{1}{4}(I(D_{\phi}^{x})+ I(D_{\phi}^{y})+I(D_{\phi'}^{x})+ 
I(D_{\phi'}^{y}))$. The resulting 
information curves 
in the range $\phi\in[0,\pi/4]$
are shown in Fig. 2 and 3.

\section{Conclusion}
When looking at the curves in figure 2, one would conclude that if
Eve does not possess a quantum memory, then the best that she can do is to
resolve to intercept/resend eavesdropping. However, one should be careful
about drawing conclusions from figure 2 alone. It should be remembered 
that in
order to be able to draw the curve for the intercept/resend eavesdropping
it was assumed that Eve was intercepting only a fraction of the qubits,
hence the information that Eve possess in this case should be viewed as an
average information on the full key. In the
intercept/resend strategy Eve's handlebar for controlling the disturbance 
$D_B$
is by intercepting only a fraction of the qubits, naturally she has no 
information on the qubits she doesn't eavesdrop on, whereas on the qubits 
that she eavesdrop she will actually have a lot of information.  
On the other hand, 
when
Eve uses an ancilla, she is interacting with each single qubit and the
disturbance $D_B$ is determined by the strength of her interaction (which
is assumed to be the same for all the qubits).

However, a couple of conclusions can be made: when Eve performs 
intercept/~resend eavesdropping on all the transmitted qubits she 
introduces a disturbance $D_{B}=1/4$ independently of the measurement she 
has chosen. Considering now the eavesdropping strategy with the ancilla: 
in order to get the same amount of information as for the 
intercept/resent eavesdropping intercepting all qubits, Eve will introduce 
a disturbance which is twice as big, namely $1/2$. At first it may seem 
curious that Eve by performing two so different eavesdropping strategies 
will end up with the same amount of information. However, is should be 
remembered that the optimal eavesdropping strategy is symmetric with 
respect to Eve and Bob and that when the disturbance is $D_B=1/2$ it 
corresponds to interchanging Eve and Bob. This basically means that Eve 
keeps the qubit sent by Alice and prepares a new qubit at random in one of 
the four states and sends it to Bob. Which makes it immediately clear that 
in this situation Eve's information corresponds to the information she 
would have gotten in the intercept/resend eavesdropping --- but since she 
sends Bob one of the four states at random, she obviously introduce a much 
higher disturbance.

As of today, the scenario we have considered here is actually quite 
realistic, since there is still a long way before having quantum 
memories which will allow an eavesdropper to store her ancilla for the 
required amount of time. The eavesdropper will then be forced to perform 
her measurement immediately, and as we have just seen in this situation a 
simple intercept/resend eavesdropping strategy is actually what will 
optimize her information.

\section*{Acknowledgment}
This work has been supported by EC under project SECOQC (contract n.
IST-2003-506813)

\end{document}